\title{Bayesian Hierarchical Spatial Regression Models for Spatial Data in the Presence of Missing Covariates with Applications}

\author{Zhihua Ma~~~~Guanyu Hu~~~~Ming-Hui Chen}

\date{}
\documentclass[12pt]{article}

\usepackage{newtxmath,newtxtext}

\usepackage[margin=1in, lines=25]{geometry}
\usepackage{graphicx}
\usepackage{subfigure}
\usepackage{arydshln}
\usepackage{caption,color,bm}
\RequirePackage[colorlinks,citecolor=cyan,urlcolor=cyan]{hyperref}

\usepackage{xr-hyper}
\usepackage{mathrsfs}
\usepackage{float}
\usepackage{amsmath,amssymb}
\usepackage{booktabs}
\usepackage{multirow}
\usepackage{threeparttable}
\usepackage{listings}
\usepackage{float}
\usepackage[T1]{fontenc}
\usepackage{newtxmath,newtxtext}

\usepackage{achemso}
\setkeys{acs}{articletitle = true}

\begin{document}
\maketitle
\begin{abstract}
In many applications, survey data are collected from different survey centers
in different regions.
It happens that in some circumstances, response variables are completely observed while the covariates have missing values.
In this paper, we propose a joint spatial regression model for the response
variable and missing covariates
via a sequence of one-dimensional conditional spatial regression models.
We further construct a joint spatial model for missing covariate data mechanisms.
The properties of the proposed models are examined and a Markov chain Monte
Carlo sampling
algorithm is used to sample from the posterior distribution.
In addition, the Bayesian model comparison criteria,  the modified Deviance
Information Criterion (mDIC)
and the modified Logarithm of the Pseudo-Marginal Likelihood (mLPML), are
developed to assess the fit of spatial regression models for spatial data.
Extensive simulation studies are carried out to examine empirical
performance of the proposed methods.
We further apply the proposed methodology to analyze a real data set from a
Chinese Health and Nutrition Survey (CHNS)
conducted in 2011.

\vspace{0.2cm}	
\end{abstract}

\section{Introduction}
Household income is a very important measurement of the development of one
region's economy. It is of great practical interest to examine the effects of
covariates
on the household income. Since household income data are always collected from
different survey centers in different regions, there are two challenges when
analyzing household income data. For geographically distributed data,
it is not desirable to fit a traditional regression model because the traditional
regression model does not account for the spatial dependence among different
regions. As a result, the first challenge for spatially dependent data such as
household incomes from different regions is to build a suitable regression
model. From \citet{banerjee2014hierarchical} and \citet{cressie2015statistics},
there are different approaches for modelling spatially dependent
data, such as the conditional autoregressive model (CAR), the simultaneous
autoregressive model (SAR), and the linear regression model with spatial random
effects. For the areal data, CAR and SAR are two widely used models. The study region is partitioned into a finite number of areal units with well-defined boundaries \citep{banerjee2014hierarchical}. The spatial correlation structure depends on adjacency matrix of subareas. The CAR model is appropriate for situations with the first order dependency or a relatively local spatial autocorrelation, which assumes that a particular area is influenced by its neighbors. However, the SAR model is more suitable where there is the second order dependency or a more global spatial autocorrelation. The locations of the point reference data vary continuously over the study region. The spatial correlation structure depends on the distance between the locations. The most popular model for point reference data is the regression model with Gaussian spatial random effects \citep{cressie1993statistics}.
Another challenge for analyzing such kind of data is that there exist some
missing covariates. Household income data are collected from surveys, so it is
common for us to get incomplete data for some covariates. There is rich
literature on building regression models with missing covariates.
\citet{zhao1996regression} used estimating equations for regression analysis in
the presence of missing observations on one covariate.
\citet{ibrahim2002bayesian} proposed methods for Bayesian inference of
regression models with missing covariates. However, no existing literature
deals with spatial data and missing covariates simultaneously.
\citet{seshadri2018statistics} proposed a spatial averaging approach for modelling spatial response data only.
\citet{bae2018missing}, \citet{xue2017fill} and
\citet{collins2017spatiotemporal} also proposed some approaches for
dealing with spatial missing data. However, they did not consider missing data
model in their approaches. Besides, spatial random effects are not commonly used in missing covariate models to take account of spatial effects.
Recently, \citet{grantham2018spatial} built a joint hierarchical model for PM 2.5 and aerosol optical depth (AOD).
To deal with missingness of AOD in spatial regression model, they assume informative missingness of AOD and build spatial regression
model for AOD to interpolate AOD.

In this paper, we develop a Bayesian spatial regression model to deal with the
spatially dependent data with missing covariates using the idea from
\citet{ibrahim2002bayesian}. We assume that the missing covariates are spatially dependent
and build hierarchical spatial regression models for both the response variable and
missing covariates. Furthermore, we propose the modified Deviance Information Criterion (mDIC)
and the modified Logarithm of the Pseudo-Marginal Likelihood (mLPML).
One of the main focus of this paper is on the examination of the impact of
spatial effects in the missing covariates models on the spatial response model.
Our proposed mDIC and mLPML criteria allow us to assess the fit of the spatial
response data under covariates models with or without
spatial effects. We further conduct extensive simulation studies to examine the empirical performance of the proposed criteria.
Such investigation and assessment have not been carried out in the literature based on our best knowledge.

The remainder of this paper is organized as follows. In Section \ref{datasection}, the data from
Chinese Health and Nutrition Survey (CHNS) 2011 are introduced as a motivating
example. In Section \ref{methodsection}, we develop the spatial regression model for the
response variable, the model for missing covariates with spatial random
effects, and the model for the missing data mechanism. Furthermore, Bayesian
model assessment criteria including mDIC and mLPML are used for model
comparison. An extensive simulation study is conducted in Section \ref{simsection} to investigate empirical performance of the models proposed in Section \ref{methodsection}.
In Section \ref{appsection}, the proposed method is
employed to analyze the real data set of CHNS 2011. Finally, we conclude the paper with a brief discussion in Section \ref{discussion}.

\section{Motivating Example} \label{datasection}
Chinese Health and Nutrition Survey (CHNS), a project collaborated by the Carolina Population Center at the
University of North Carolina and the National Institute for Nutrition and
Health at the Chinese Center for Disease Control and Prevention, aims to
examine the relationship between the social and economic transformation of Chinese
society and the health and nutritional status of its population. As a
geographically distributed data set, CHNS 2011 collected individual-,
household- and community-specific information from 12 provinces in China. In
this paper, household income from 12 provinces is selected as the spatial
response variable, and the aim is to explore the spatial effects and the
factors that may have impacts on this variable of interest.

\subsection{Data Description}
The data were collected from 12 provinces in China with a total sample size
of 4346. Household income (\textit{hincome}) is the response variable.
Individual-level covariates include wage of head of the household (\textit{indwage}),
age of head of the household (\textit{age}), proportion of urban area (\textit{urban}), number of hours
worked last year (\textit{WThour}), family size (\textit{hhsize}) and GDP per capita of the
province (\textit{GDP}). The units of \textit{hincome}, \textit{indwage} and \textit{GDP} are CNY.

The sample sizes in different provinces as well as the summary information of the variables are shown in Table
\ref{table5}.

\begin{table}[tbp]
	\centering
	\caption{Sample size and summary information of the variables in each province} \label{table5}
	\begin{tabular}{cccccccc}
		\hline
		& &Beijing & Liaoning & Heilongjiang & Shanghai & Jiangsu & Shandong \\
		\hline
		\multicolumn{2}{c}{Sample size}&415 & 395 & 396 & 424 & 412 & 399 \\
		\multirow{2}{*}{\textit{hincome}} & mean & 75599.23& 49426.97 & 46861.01 &87455.34 &61393.95&40999.05\\
		& sd & 49926.75 & 47862.42 & 44386.13 &68695.15&  43495.38& 40926.51\\
		\multirow{2}{*}{\textit{indwage}} & mean & 41029.76& 25021.40& 29590.38& 41829.25 &20894.34 &20769.75 \\
		& sd & 41730.44 &  25584.54 &  40866.63 &45704.46&20348.74& 24941.24\\
		\multirow{2}{*}{\textit{age}} & mean & 49.30 &56.40   & 51.15 &56.28 & 59.31 &56.01     \\
		& sd & 13.13 &  11.88 &    11.42 &   11.68&    11.83&11.38\\
		\multirow{2}{*}{\textit{urban}} & proportion & 0.86 &0.30& 0.37 & 0.83 & 0.33&0.29\\
		& sd &  0.34 & 0.46 &  0.48  & 0.38& 0.47 &0.46 \\
		\multirow{2}{*}{\textit{WThour}} & mean & 44.21 &38.15 & 27.78  &  41.43 &35.86  &  43.09 \\
		& sd & 12.22  &    24.77  &  24.50&    8.77  &   22.49&    18.10\\
		\multirow{2}{*}{\textit{hhsize}} & mean &2.80 &2.85   &2.62 &  3.20& 3.22&3.01  \\
		& sd &0.84 &   1.16  &   1.01 &      1.10&     1.51&   1.33\\
		\hline
		&& Henan & Hubei & Hunan & Guangxi & Guizhou & Chongqing \\
		\hline
		\multicolumn{2}{c}{Sample size}&298 & 337 & 244 & 360 & 339 & 327\\
		\multirow{2}{*}{\textit{hincome}} & mean & 36782.92 &50417.49 &48163.62 &37022.83 & 45388.52 &41770.31  \\
		& sd &42655.65& 57341.40&43458.89&33374.35&52696.84& 39894.29\\
		\multirow{2}{*}{\textit{indwage}} & mean & 16022.50&21769.90  &27191.26 &12122.19 &22694.39 &25977.72\\
		& sd & 24142.06 &28241.78&29676.95&14334.89&39639.95& 34609.48\\
		\multirow{2}{*}{\textit{age}} & mean &53.96& 54.67   & 53.39    &55.27   &56.21    &52.48   \\
		& sd &   12.18& 10.38&12.44&  12.43 & 12.58&   11.52  \\
		\multirow{2}{*}{\textit{urban}} & proportion & 0.37 &0.33  &  0.41  &0.29  &0.33 &  0.54      \\
		& sd &    0.48 &    0.47 &    0.49&    0.45&    0.47 &0.50\\
		\multirow{2}{*}{\textit{WThour}} & mean &   37.12&38.23 &39.31 & 36.32   &32.07    & 38.85  \\
		& sd &  23.11&   18.53&   17.01&  21.36&18.95&   18.97\\
		\multirow{2}{*}{\textit{hhsize}} & mean &3.67 & 3.27     &3.25   &  4.14   & 3.43 & 3.20     \\
		& sd &   1.49  & 1.48  &   1.35&   1.76 &    1.43 &1.11 \\
		\hline
	\end{tabular}
\end{table}

Among these covariates, \textit{indwage} and \textit{WThour} have
missing values. The average percentages with only \textit{indwage} or \textit{WThour} missing are 22.50\% and 5.34\%, respectively, while the average percentage with both \textit{indwage} and \textit{WThour} missing is 22.30\%. A summary of the missing patterns of these two covariates are
given in Table \ref{table6}.

\begin{table}[tbp]
	\centering
	\caption{Missing percentages in each province} \label{table6}
	\makebox[\linewidth][c]{
		\begin{tabular}{ccccccc}
			\hline
			& Beijing & Liaoning & Heilongjiang & Shanghai & Jiangsu &
			Shandong \\
			\hline
			missing \textit{indwage} only & 1.20\% & 25.57\%  & 41.92\%  & 0.94\% & 18.20\% & 21.55\%  \\
			missing \textit{WThour} only & 4.82\% & 3.04\% & 4.55\% & 3.54\% & 7.52\% & 8.77\%\\
			missing \textit{indwage} and \textit{WThour} & 30.12\% & 29.37\%  & 12.37\%  & 43.40\% & 19.90\% & 26.32\%  \\
			\hline
			& Henan & Hubei & Hunan & Guangxi & Guizhou & Chongqing  \\
			\hline
			missing \textit{indwage} only & 22.48\% & 32.94\% & 23.77\% & 28.33\% & 31.86\% & 29.05\%\\
			missing \textit{WThour} only & 5.37\% & 5.93\% & 4.10\% & 6.94\% & 2.95\% & 6.12\%\\
			missing \textit{indwage} and \textit{WThour}  & 12.75\% & 15.13\% & 19.67\% & 13.89\% & 17.40\% & 18.96\%\\
			\hline
		\end{tabular}
	}
\end{table}

\subsection{Spatial Structure}
\begin{figure}[t!]
    	\centering
	\includegraphics[scale=0.8]{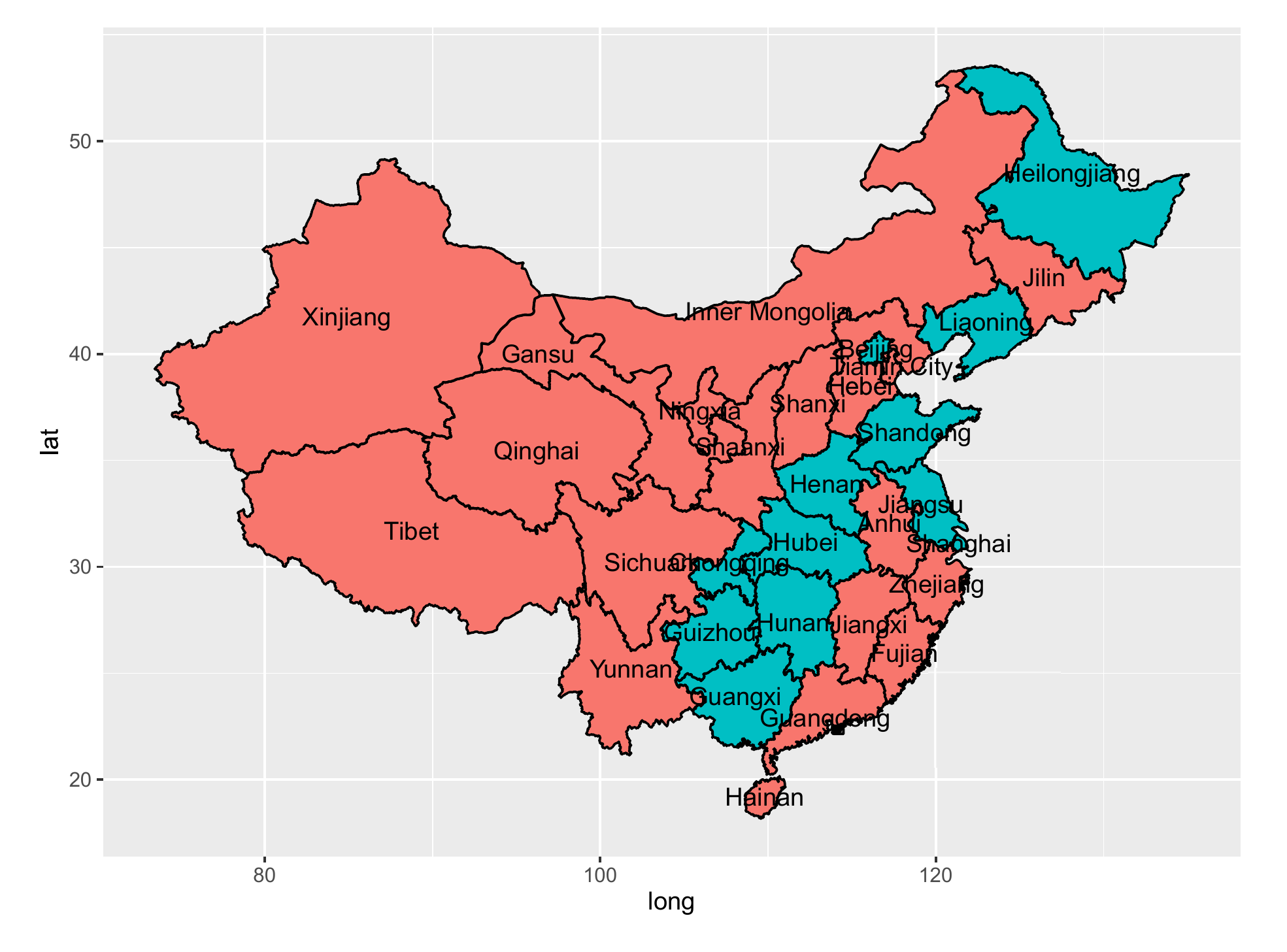}
	\caption{China Map (Blue indicates the province which are included in the study)}
	\label{map}
\end{figure}

In the CHNS 2011 data set, we do not have survey data for all the provinces in
China. Also, the provinces included in this data set are not always neighbored with each other.
Thus, we treat the CHNS 2011 data as point-referenced data such that the spatial dependence
can be possibly and reasonably captured by the distance between two provinces especially when they are away from each other.
The centroid latitudes and longitudes of these 12 provinces are given in Table \ref{coordinates}. Figure \ref{map}
shows the map of mainland China. The provinces which are included in our study are marked in blue color.
\begin{table}[tbp]
	\centering
	\caption{Centroid Coordinates of each province} \label{coordinates}
	\makebox[\linewidth][c]{
		\begin{tabular}{ccccccc}
			\hline
			Province & Beijing & Liaoning & Heilongjiang & Shanghai & Jiangsu & Shandong \\
			\hline
			Longitude  & 116.4107 & 122.6090 & 127.7824 & 121.4037 & 119.4554 & 118.1490  \\
			Latitude & 40.1849 & 41.3037 & 47.8415 & 31.0846 & 32.9732 & 36.3512\\
			\hline
			Province & Henan & Hubei & Hunan & Guangxi & Guizhou & Chongqing  \\
			\hline
			Longitude & 113.6136 & 112.2691 & 111.7083 & 108.7872 & 106.8738 & 107.8748\\
			Latitude & 33.8826 & 30.9760 & 27.6069 & 23.8279 & 26.8152 & 30.0587\\
			\hline
		\end{tabular}
	}
\end{table}

Using the coordinates of 12 provinces, we can easily calculate the distance
between two provinces. These distances are useful to construct covariance matrices of the spatial random effects in Section \ref{appsection} below.

\section{Methodology} \label{methodsection}

In this section, a spatial regression model with missing covariates is built hierarchically.  A Gaussian spatial regression model for the response variable is built, after which, missing covariate distributions are built to take account of the
missing covariates and covariate-specific spatial effects. In addition, a model capturing the
missing data mechanism is also built. After introducing the model construction, posterior inference procedure and model assessment are presented.

\subsection{The Spatial Regression Model for Responses} \label{resmodelsection}

Suppose, we consider $S$ locations and $N_s$ observations at location $s$
$(s=1,\cdots,S)$. The spatial response variable at location $s$ is denoted by
$Y(s)=(Y_1(s),\cdots,Y_{N_s}(s))^\prime$. A Gaussian stationary spatial process model
is built for the spatial response variable. The general Gaussian stationary spatial process model can be written
as in, for example, \citet{cressie1993statistics}:
\begin{equation}
Y(s)=\bm{X}(s)^{'}\bm{\beta}+\sigma_y W_y(s)\bm{1}_{N_s}+\epsilon(s),
\label{spatial regression}
\end{equation}
where $\bm{X}(s) = (\bm{1}_{N_s}, X_{1}(s),\dots,X_p(s))'$ is a $(p+1) \times N_s$ matrix, $p$ is the number of covariates, $\bm{1}_{N_s}$ is the $N_s$-dimensional vector with $1$s,
$X_k(s)=(X_{k1}(s),\cdots,X_{kN_s}(s))^\prime$ is an $N_s$-dimensional vector of covariates, and $\bm{\beta} = (\beta_0, \beta_1, \cdots, \beta_p)'$ is a $(p+1)$ dimensional vector of corresponding regression coefficients. The spatial random effect $W_y(s)$ is a second-order stationary mean-zero process. To be more specific,
$W_y(s)$ conforms that $\text{E}(W_y(s))=0$, $\text{Var}(W_y(s))=1$,
and $\text{Cov}(W_y(s), W_y(s'))=\rho(s,s')$, where $\rho(\cdot)$ is
a valid two-dimensional correlation function.
$\epsilon(s)$ is the white noise process such that $\epsilon(s) \sim \text{MVN}(\bm{0}_{N_s},\tau_y^{-1}\bm{I}_{N_s})$, where  ``MVN'' represents the multivariate normal distribution, $\bm{I}_{N_s}$ is the $N_s \times N_s$ identity matrix, and
$\text{Cov}(\epsilon(s),\epsilon(s'))=0$ for $s \neq s'$.
According to
\eqref{spatial regression}, the following spatial hierarchical model is built:
\begin{align}
Y(s) | W_y(s), \bm{X}(s), \bm{\beta}, \sigma_y, \tau_y &\sim \text{MVN}(\bm{X}(s)^{'}
\bm{\beta}+\sigma_y W_y(s)\bm{1}_{N_s}, \tau_y^{-1}\bm{I}_{N_s}),\qquad s=1,2,\ldots,S, \label{responsemodely}\\
\bm{W}_y | \lambda_y &\sim \text{MVN}(\bm{0},H(\lambda_y)),
\label{responsemodelw}
\end{align}
where $\bm{W}_y=(W_y(1),\cdots, W_y(S))'$ is the response-specific spatial random effect,
$H(\lambda_y)$ is a spatial correlation matrix based
on distance and parameter $\lambda_y$. For the exponential spatial correlation
kernel, the $(s,s^{'})$th entry of the correlation matrix is $\exp(-\lambda_y
d_{ss^{'}})$, where $d_{ss^{'}}$ is the Euclidian distance between location $s$
and location $s^{'}$, and $\lambda_y$ is the range parameter for spatial correlation. A small
value of $\lambda_y$ means a strong spatial correlation, and a large value of  $\lambda_y$ means a weak
spatial correlation.

%


\subsection{The Spatial Regression Models for Missing Covariates}

For survey data, it is common that the data for some covariates are not completely observed. For example, $X_k(s)$ is the $k$th covariate at location $s$ and has $N_s$ observations. If there are any missing values among those $N_s$ observations, i.e. if any one of the elements of $(X_{k1}(s), X_{k2}(s), \cdots, X_{kN_s}(s))$ is missing, $\bm{X}_k$ is defined as a missing covariate at location $s$.  For the CHNS 2011 dataset discussed in Section \ref{datasection}, two missing covariates exist at all locations. Therefore, in this section, we assume that for all locations, among the $p$ covariates, the first $q$ $(q \le p)$ covariates are missing covariates.

In the presence of missing covariates, a joint model for the missing covariates
should be specified to take account of the uncertainty resulting from the
missing values in the covariates. To be specific, for the $i$th observation at location $s$, the
corresponding $q$-dimensional missing covariate vector is
$\bm{X}_i^{mis}(s)=(X_{1i}(s),\cdots,X_{qi}(s))'$, while the $(p-q)$-dimensional
complete covariate vector is denoted by
$\bm{X}_i^{obs}(s)=(X_{q+1,i}(s),\cdots,X_{pi}(s))'$. For missing covariate data, it
is crucial to specify a model for the missing covariates $\bm{X}_i^{mis}(s)$. Given the spatial random effects, we assume $\bm{X}_i^{mis}(s)$, $i=1,2,\ldots,N_s$, are conditionally independent. In
general settings, \citet{lipsitz1996conditional} and \citet{ibrahim1999missing}
specified the missing covariate distribution through a series of
one-dimensional conditional distributions.
In our case, since the covariates are also spatially distributed,
covariate-specific spatial effects are also taken into account in the missing
covariate model. We extend their model 
as

\begin{align}
f(X_{1i}(s), \cdots,
X_{qi}(s)&|\bm{X}_i^{obs}(s),\bm{W}_x(s),\bm{\alpha},\bm{\sigma}_x,\bm{\tau}_x) =
\nonumber \\ &f(X_{qi}(s)|X_{1i}(s),\cdots,
X_{q-1,i}(s),\bm{X}_i^{obs}(s),W_{x_q}(s),\bm{\alpha}_q,\sigma_{x_q},\tau_{x_q})
\nonumber \\
& \times
f(X_{q-1,i}(s)|X_{1i}(s),\cdots,X_{q-2,i}(s),\bm{X}_{i}^{obs}(s),W_{x_{q-1}}(s),\bm{\alpha
}_{q-1},\sigma_{x_{q-1}},\tau_{x_{q-1}})  \nonumber\\
&\times \cdots\times
f(X_{1i}(s)|\bm{X}_i^{obs}(s),W_{x_1}(s),\bm{\alpha}_1,\sigma_{x_1},\tau_{x_1}),
\label{missingcov2}
\end{align}
where $\bm{W}_x(s) = (W_{x_1}(s), \cdots, W_{x_q}(s))'$, $W_{x_\ell}(s)$ represents the spatial effect of covariate $X_{\ell,i} (\ell=1,\cdots,q)$ at
location $s$, $\bm{\sigma}_x=(\sigma_{x_1},\cdots,\sigma_{x_q})^{'}$ is a vector of the standard deviations of the spatially structured random errors of the covariates,
$\bm{\tau}_x=(\tau_{x_1},\cdots,\tau_{x_q})^{'}$ is a vector of the precisions of the independent random errors of the covariates, and the coefficients associated to the covariates are $\bm{\alpha}=(\bm{\alpha}_1,\cdots,\bm{\alpha}_q)'$ with $\bm{\alpha}_\ell$ being the indexing parameter vector for the $\ell$th conditional distribution. For the covariate-specific spatial random effects $\bm{W}_{x_\ell}=(W_{x_\ell}(1),\cdots,W_{x_\ell}(S))^{'}$, a multivariate normal
distribution similar to (\ref{responsemodelw}) can be assumed. As in \citet{grund2016multiple},
here we assume the spatial random effects, $\bm{W}_{x_\ell}$'s, of the missing covariates and $W_y(s)$ of the response variable are independent.
This assumption is reasonable since
$\bm{W}_{x_\ell}$ captures spatial dependence of the covariate $x_{\ell}(s)$, 
$W_y(s)$ captures spatial dependence of the response variable,
and the dependence between the response variable and the covariates is induced by the spatial regression model in
(\ref{responsemodelw}).

There are many possibilities in ($\ref{missingcov2}$), especially
when $q$ is large. \citet{chen2001maximum} gave some guidelines for specifying
the sequence of one-dimensional conditional distributions. When the missing
covariates are categorical, logistic regression for the conditional missing
covariate distribution can be specified. Probit or complementary log-log links
are also suitable to model categorical covariates. Ordinal regression models
can be employed to model missing ordinal covariates. For count variables, we
can model them via Poisson regression. And for continuous variables, normal regression, log-normal regression, and exponential regression can be considered.

In our extended model, covariate-specific spatial effects are considered in the
model additionally. Conditional on the spatial effects, missing covariates can be modeled according to the above strategy.
And for the spatial effects, the same stationary process structure in
(\ref{responsemodelw}) can be used. In the motivating example, there are $q=2$
missing continuous covariates, and the spatial regression model for these two
missing covariates can be written, for $i=1,2,\cdots,N_s$ and $s=1,2,\cdots,S$, as
\begin{gather*}
X_{2i}(s)|X_{1i}(s), \bm{X}_i^{obs}(s), W_{x_2}(s), \bm{\alpha}_2,
\sigma_{x_2},\tau_{x_2} \sim N(\bm{X}^{(-2)}_{i}(s)' \bm{\alpha}_2  +
\sigma_{x_2}W_{x_2}(s),\tau_{x_2}^{-1}),\\
X_{1i}(s)|\bm{X}_i^{obs}(s),W_{x_1}(s), \bm{\alpha}_1,\sigma_{x_1},\tau_{x_1} \sim
N(\bm{X}_i^{obs}(s)^{'} \bm{\alpha}_1 + \sigma_{x_1} W_{x_1}(s), \tau_{x_1}^{-1}),\\
\bm{W}_{x_2} | \lambda_{x_2} \sim \text{MVN}(\bm{0},H(\lambda_{x_2})), \hspace{0.2cm}\bm{W}_{x_1} | \lambda_{x_1} \sim \text{MVN}(\bm{0},H(\lambda_{x_1})),
\label{missingcov3}
\end{gather*}
where $\bm{X}^{(-2)}_{i}(s)=(X_{1i}(s),(\bm{X}_i^{obs}(s))^\prime)^\prime$ denotes a vector of the other covariates except $X_{2i}(s)$, $\bm{\alpha}_1$ and $\bm{\alpha}_2$ are the indexing parameter vectors for the distributions of $X_{1i}(s)$ and $X_{2i}(s)$, respectively, $\tau_{x_1}$ and $\tau_{x_2}$ are the precision parameters of $X_{1i}(s)$ and $X_{2i}(s)$, $\sigma_{x_1}$ and $\sigma_{x_2}$ are the standard deviations of $\bm{W}_{x_1}$ and $\bm{W}_{x_2}$, and $\lambda_{x_1}$ and $\lambda_{x_2}$ are the corresponding range parameters for spatial correlations of $\bm{W}_{x_1}$ and $\bm{W}_{x_2}$, which are different than $\lambda_y$ defined in (\ref{responsemodelw}).

\subsection{Models for Missing Data Mechanism}
Assuming a corresponding missing indicator for each missing covariate, for observation $i$ we
have the $q$-dimensional missing indicator vector
$\bm{R}_i(s)=(R_{1i}(s),\cdots,R_{qi}(s))'$ with $R_{\ell i}(s)=1$ if $X_{\ell i}(s)$ is observed
and $R_{\ell i}(s)=0$ if $X_{\ell i}(s)$ is missing ($\ell=1,\cdots,q$). The joint
distribution of $R_{\ell i}(s)$ can also be written as the form of a product of
one-dimensional conditional distributions, that is 
\begin{align}
\label{eq:rmodel}
f(R_{1i}(s),\cdots,R_{qi}(s)|\bm{X}_i(s),Y_i(s),\bm{\phi}) =& f(R_{q i}(s)|
R_{1 i}(s),\cdots,R_{q-1,i}(s), \bm{X}_i(s),Y_i(s),\bm{\phi}_q)\notag\\
&\times \cdots \times f(R_{1i}(s)|\bm{X}_i(s),Y_i(s),\bm{\phi}_1)
\end{align}
for $i=1,2,\cdots,N_s$ and $s=1,2,\cdots,S$, where $\bm{\phi}=(\bm{\phi}_1,\cdots,\bm{\phi}_q)'$ parameterizes the
missingness mechanism model with $\bm{\phi}_\ell$ as a vector of indexing
parameters for the $\ell$th conditional distribution. For each one-dimensional
conditional distributions of these binary missing indicators, it is common to
build a logistic regression model for each of them.

In the missing data literature, missing data mechanism can be categorized as
missing completely at random (MCAR), missing at random (MAR) or missing not at random (MNAR) \citep{rubin1976inference}. When missingness does not depend on the covariates that are missing or observed, then the missing data mechanism is termed as MCAR. When missingness
depends only on the observed covariates but not on the missing ones, the missing data mechanism is MAR. When neither MCAR nor MAR holds, the missing data mechanism is termed as MNAR.

For simplicity, in our case we assume that the missing data mechanism is MAR,
which means that the missing data does not depend on the missing covariates. For $q=2$ missing covariates, the joint distribution of the missing
indicators is written as
\begin{gather*}
f(R_{1i}(s), R_{2i}(s)|\bm{X}_i^{obs}(s),Y_i(s),\bm{\phi}) =
f(R_{2i}(s)|R_{1i}(s),\bm{X}_i^{obs}(s),Y_i(s),\bm{\phi}_2) \times f(R_{1i}(s)|\bm{X}_i^{obs}(s),Y_i(s),\bm{\phi}_1), \\
R_{2i}(s)|R_{1i}(s),\bm{X}_i^{obs}(s),Y_i(s),\bm{\phi}_2 \sim \text{Bernoulli}(p_{2i}(s)),\\
R_{1i}(s)|\bm{X}_i^{obs}(s),Y_i(s),\bm{\phi}_1 \sim \text{Bernoulli}(p_{1i}(s)),\\
\text{logit}(p_{2i}(s)) = \text{log}(p_{2i}(s)/(1-p_{2i}(s))) = (\bm{X}_i^{obs}(s)', Y_i(s))'\bm{\phi}_2,\\
\text{logit}(p_{1i}(s)) = (\bm{X}_i^{obs}(s)', Y_i(s))'\bm{\phi}_1.
\label{missingR}
\end{gather*}

\subsection{Inference Procedure} \label{priorsection}
For the unknown parameters $\bm{\theta} = \{\bm{\beta}, \sigma_y, \tau_y, \lambda_y, \bm{\alpha},\bm{\sigma}_x,\bm{\tau}_x, \bm{\lambda}_x, \bm{\phi}\}$, where $\bm{\lambda}_x = \{ \lambda_{x_{\ell}} \}_{\ell = 1}^q$, we assume that they are independent a priori. For $\ell=1,\cdots,q$, the following
prior distributions are assigned: $\beta_k \sim N(0,\psi_{\beta_k}^{-1})$,
for $k=0,\cdots,p$;
$\tau_y^{-1} \sim \text{IG}(a_y,b_y)$;
$\sigma_y^2 \sim \text{half-Normal}(0, \psi_{\sigma_y}^{-1})$;
$\lambda_y \sim \text{log-Normal}(0, \psi_{\lambda_y}^{-1})$;
$\alpha_{\ell k} \sim N(0, \psi_{\alpha_{\ell k}}^{-1})$, for $k=0,\cdots,m_\ell$;
$\sigma_{x_\ell}^2 \sim \text{half-Normal}(0, \psi_{\sigma_{x_\ell}}^{-1})$;
$\tau_{x_{\ell}}^{-1} \sim \text{IG}(a_{x_\ell},b_{x_\ell})$;
$\lambda_{x_\ell} \sim \text{log-Normal}(0,$ $\psi_{\lambda_{x_\ell}}^{-1})$;
and $\phi_{\ell k} \sim N(0,\psi_{\phi_{\ell k}}^{-1})$, for $k=0,\cdots,m_\ell^{'}$,
where $m_\ell$ and $m_\ell^{'}$ are the dimensions of covariates in the missing
covariate model and missing data mechanism model of the $\ell$th missing
covariate, respectively.

Note that $\psi_{\beta_k}$, $a_y$, $b_y$, $\psi_{\sigma_y}$,
$\psi_{\lambda_y}$, $\psi_{\alpha_{\ell k}}$, $\psi_{\sigma_{x_\ell}}$, $a_{x_\ell}$,
$b_{x_\ell}$, $\psi_{\lambda_{x_\ell}}$, and $\psi_{\phi_{\ell k}}$ are prespecified
hyperparameters. In this article, we use
$\psi_{\beta_k}=\psi_{\alpha_{\ell k}}=\psi_{\sigma_y}=\psi_{\sigma_{x_\ell}}=\psi_
{\phi_{\ell k}}=0.001$, $a_y=b_y=a_{x_\ell}=b_{x_\ell}=0.001,$
and $\psi_{\lambda_y}=\psi_{\lambda_{x_\ell}}=1$, which lead to non-informative priors.
With the above prior distributions, the posterior distribution of these unknown
parameters based on the observed data $D_{obs}$ $= \{\bm{Y}, \bm{X}^{obs} \} $ with $\bm{X}^{obs} = \{ \bm{X}^{obs}(s) \}_{s=1}^S$ is given by
\begin{equation}
\begin{split}
\pi(\bm{\theta} | D_{obs}) &\propto L(\bm{\theta}|D_{obs}) \pi(\bm{\theta})  \\
& \propto \bigg[\int \prod_{s=1}^S\Big( f(\bm{Y}(s) | W_y(s), \bm{X}(s), \bm{\beta}, \sigma_y, \tau_y )\\
&\prod_{i=1}^{N_s}\int f(\bm{X}_i^{mis}(s)|\bm{X}_i^{obs}(s),\bm{W}_x(s),\bm{\alpha},\bm{\sigma}_x, \bm{\tau}_x)f(\bm{R}_i(s)|\bm{X}_i(s), Y_i(s), \bm{\phi})d\bm{X}_i^{mis}(s)\Big)\\
& \hspace{0.8in}\times f(\bm{W}_y|\lambda_y) f(\bm{W}_x|\bm{\lambda}_x)
d\bm{W}_y d\bm{W}_x \bigg]  \pi(\bm{\theta}),	
\end{split}
\label{eq:inference}
\end{equation}
where $f(\bm{Y}(s) | W_y(s), \bm{X}(s), \bm{\beta}, \sigma_y, \tau_y)$ refers to the
spatial regression model for the response variable in (\ref{responsemodely}), $f(\bm{X}_i^{mis}(s)|\bm{X}_i^{obs}(s),\bm{W}_x(s),\bm{\alpha},\bm{\sigma}_x, \bm{\tau}_x)$ is defined in \eqref{missingcov2}, and $f(\bm{R}_i(s)|\bm{X}_i(s), Y_i(s), \bm{\phi})$ is defined in \eqref{eq:rmodel} . In equation \eqref{eq:inference}, $f(\bm{W}_y|\lambda_y)$ and $f(\bm{W}_x|\bm{\lambda}_x)=\prod^q_{\ell=1}f(\bm{W}_{x_\ell}|\bm{\lambda}_{x_\ell})$
refer to the distributions of $\bm{W}_y$ and $\bm{W}_{x_\ell}$'s, respectively, $d\bm{W}_x= \prod^q_{\ell=1} d \bm{W}_{x_\ell}$,
and
$\pi(\bm{\theta})$ denotes the joint prior distribution of the unknown
parameters. When a MAR missing data mechanism is assumed, the model for the missing data mechanism does not need to enter the posterior distribution.

The analytical form of the posterior distribution of $\bm{\theta}$ is
unavailable. Therefore, we carry out the posterior inference using the Markov
chain Monte Carlo (MCMC) sampling algorithm to sample from the posterior
distribution. Instead of sampling from the posterior distributions of the
unknown parameters directly, MCMC samples from the full conditional
distributions of the parameters with the remaining variables fixed to their
current values are obtained. In this way, we can conduct inferences of the proposed model.
In our case, spatial random effects are also regarded as unknown parameters,
and then the algorithm samples these parameters in turn from their
corresponding full conditional distributions.

\subsection{Model Assessment}  \label{modelassessmentsection}
Within the Bayesian framework, the Deviance Information Criterion (DIC)
\citep{spiegelhalter2002bayesian} and the Logarithm of the Pseudo-Marginal
Likelihood (LPML) \citep{ibrahim2013bayesian} are two well-known Bayesian
criteria for model comparison.

Since our main objective is to assess the fit of the spatial regression model for the response, we specify the following deviance function:
\begin{equation}
\begin{split}
		&\text{Dev}(\bm{W}_y,\bm{X},\bm{\beta},\sigma_y,\tau_y)=-2\sum_{s=1}^S
\text{log} f(\bm{Y}(s) | W_y(s), \bm{X}(s), \bm{\beta}, \sigma_y, \tau_y)\\
&=\sum_{s=1}^S\Big\{ \text{log}(2\pi)+2 N_s\text{log}(\tau_y)\\
&+(\bm{Y}(s) - \bm{X'(s) \beta}-\sigma_y W_y(s)\bm{1}_{N_s})' |\tau_y^2 I_{N_s}|^{-1} (\bm{Y}(s) - \bm{X(s)' \beta}-\sigma_y W_y(s)\bm{1}_{N_s}) \Big\}.
\end{split}
\label{deviance}
\end{equation}

Therefore, we define a modified DIC (mDIC) for the response model as
follows:
\begin{align}
\text{mDIC}=2 E[\text{Dev}(\bm{W}_y,\bm{X},\bm{\beta},\sigma_y,\tau_y)] -
\text{Dev}(\hat{\bm{W}}_y,\hat{\bm{X}},\hat{\bm{\beta}},\hat{\sigma}_y,\hat{\tau}_y),
\label{DIC model}
\end{align}
where $\hat{\bm{W}}_y,\hat{\bm{X}},\hat{\bm{\beta}},\hat{\sigma}_y$, and $\hat{\tau}_y$ are the posterior means of parameters and missing covariates. A
smaller value of $\text{mDIC}$ indicates a better model.

Let $D_{(-i)}(s) = \{Y_j(s): j=1,\cdots,i-1,i+1,\cdots,N_s,s=1,2,\ldots,S\}$ denote the observation
data with the $i$th subject response deleted.
Following \citet{hanson2011predictive}, we consider a modified Conditional Predictive Ordinate (mCPO) for the $i$th
subject as
\begin{align}
\text{mCPO}_i(s) = \int f(&Y_i(s)|\bm{W}_y(s),\bm{X}(s),\bm{\beta},\sigma_y,
\tau_y)  \notag \\
&\times\pi(\bm{W}_y,\bm{X},\bm{\beta},\sigma_y,
\tau_y|D_{(-i)})d(\bm{W}_y,\bm{X},\bm{\beta},\sigma_y,\tau_y),
\label{CPO def}
\end{align}
where $\pi(\bm{W}_y,\bm{X},\bm{\beta},\sigma_y, \lambda_y,\tau_y|D_{(-i)}(s)) =
\frac{\prod_{s=1}^S\prod_{j \ne i}f(Y_j(s)|\bm{W}_y(s),\bm{X}(s),\bm{\beta},\sigma_y,\tau_y) \pi(\bm{W}_y,\bm{X},\bm{\beta},\sigma_y,
\tau_y)}{c(D_{(-i)}(s))}$ and $c(D_{(-i)}(s))$ denotes the normalizing
constant. In practice, a Monte Carlo estimate of mCPO using MCMC algorithms from
the posterior distributions can be used. To be specific, letting $\bm{W}_{yt}(s),
\bm{X}_t(s), \bm{\beta}_t, \sigma_{yt}$, and $\tau_{yt}$ ($t=1,\cdots,T$)
denote a MCMC sample of unknown parameters and missing covariates from the corresponding augmented posterior distribution, a Monte Carlo
estimate of $\text{mCPO}_i^{-1}$ is given by
\begin{align}
\widehat{\text{mCPO}}_i(s)^{-1} = \frac{1}{T}\sum_{t=1}^{T}
\frac{1}{f(Y_i(s)|\bm{W}_{yt}(s),\bm{X}_t(s),\bm{\beta}_t,\sigma_{yt},\tau_{yt})}.
\end{align}
Then mLPML is given by
\begin{align}
\widehat{\text{mLPML}} =\sum_{s=1}^S\sum_{i=1}^{N_s}\text{log}(\widehat{\text{mCPO}}_i(s)).
\end{align}
Similar to the conventional LPML, a larger value of
$\text{mLPML}$ indicates a more favorable model.


\section{A Simulation Study} \label{simsection}

\subsection{Simulation Description}
In this simulation study, we randomly generated 20 locations in a space of
$[0,20]\times [0,20]$. For each location, we generated 50 observations based on
\begin{align*}
Y_i(s)=\beta_0+\beta_1X_{1i}(s)+\beta_2X_{2i}(s)+\beta_3X_{3i}(s)+\sigma_y
W_y(s)+\epsilon_i(s),
\end{align*}
where $s=1,\dots,20$, $i=1,\cdots,50$, $\epsilon_i(s)$ is i.i.d. generated from $N(0,1)$ and
$\bm{W}_y \sim \mbox{MVN}(\bm{0},H(\lambda_y))$. Covariate $X_{3i}(s)$ is
independently generated from $N(0,1)$, $X_{1i}(s)$ is generated from
$N(X_{3i}(s)+\sigma_{x_1} W_{x_1}(s),1)$, and $X_{2i}(s)$ is generated from
$N(2X_{1i}(s)+\sigma_{x_2} W_{x_2}(s),1)$,
where $\bm{W}_{x_1}\sim \text{MVN}(\bm{0}, H(\lambda_{x_1}))$, $\bm{W}_{x_2}\sim
\text{MVN}(\bm{0},H(\lambda_{x_2}))$, $\sigma_{y}=\sqrt{2}$, $\sigma_{x_1}=1$,
and $\sigma_{x_2}=\sqrt{1.5}$. For both spatial random effects, the
$(s,s')$th entry of $H(\cdot)$ is $\exp(-d_{ss'}/\lambda)$, where $d_{ss'}$
is the distance between $s$ and $s'$, $\lambda_y=3$, $\lambda_{x_1}=5$, and
$\lambda_{x_2}=4$.


Missing data for $(\bm{X}_{1}(s), \bm{X}_{2}(s))$ are generated with a missing data
mechanism that does not depend on $(\bm{X}_{1}(s), \bm{X}_{2}(s))$, leading to the
missing data to be MAR. As a result, the missing data mechanism can be ignored
when estimating the parameters. Specifically, let $R_{\ell i}(s)=1$ if $X_{\ell i}(s)$ is
observed and $R_{\ell i}(s)=0$ if $X_{\ell i}(s)$ is missing ($\ell=1,2)$. The joint
distribution of $(R_{1}(s), R_{2}(s))$ is given by
\begin{align}
f(R_{1}(s), R_{2}(s)|\bm{\phi}) = f(R_{2}(s)|R_{1}(s), \bm{\phi}_2)f(
R_{1}(s)|\bm{\phi}_1),
\end{align}
where $\bm{\phi}=(\bm{\phi}_1,
\bm{\phi}_2)$, $\bm{\phi}_1$ and $\bm{\phi}_2$ are the vectors of parameters corresponding to the distributions of $\bm{R}_1(s)$ and $\bm{R}_2(s)$, respectively. We take logistic regression models for $f(\bm{R}_{2}(s)|\bm{R}_{1}(s),$
$\bm{\phi}_2)$ and $f(\bm{R}_{1}(s)|\bm{\phi}_1)$. Thus,
\begin{align} \label{16}
f(R_{2i}(s)=1| R_{1i}(s), \bm{\phi}_2, X_{3i}(s), Y_i(s)) =
\frac{\text{exp}(\phi_{20}+\phi_{21}X_{3i}(s) + \phi_{22} Y_i(s) + \phi_{23}
	R_{1i}(s) )}{1+\text{exp}(\phi_{20}+\phi_{21}X_{3i}(s) + \phi_{22} Y_i(s) +
	\phi_{23} R_{1i}(s) )},
\end{align}
and
\begin{align} \label{17}
f(R_{1i}(s)=1| \bm{ \phi}_1, X_{3i}(s), Y_i(s)) = \frac{\text{exp}(\phi_{10}+\phi_{11}X_{3i}(s) +
	\phi_{12} Y_i(s)) }{1+\text{exp}(\phi_{10}+\phi_{11}X_{3i}(s) + \phi_{12} Y_i(s))}.
\end{align}
In (\ref{16}) and (\ref{17}), $\bm{\phi}_1 = (\phi_{10}, \phi_{11}, \phi_{12})'$ and $\bm{\phi}_2 = (\phi_{20}, \phi_{21}, \phi_{22}, \phi_{23})'$.
One hunderd simulated datasets were generated in this study. The average percentages over the 100 simulated datasets with only $\bm{X}_{1}(s)$ missing or only $\bm{X}_{2}(s)$ missing are 32.82\% and 39.27\% respectively, while the average percentage with both $\bm{X}_{1}(s)$ and $\bm{X}_{2}(s)$ missing is 28.72\%.


\subsection{Simulation Results}

According to Section \ref{methodsection}, we set up the following model $M_1$ and fix the
parameters related to $\bm{W}_y, \bm{W}_{x_2}, \bm{W}_{x_3}$ to their true
values. The
spatial regression model for the response variable is given as
\begin{gather*}
Y_i(s) \sim N(\mu_{yi}(s),\tau_y^{-1}),\hspace{2pt}\mu_{yi}(s) = \beta_0 + \beta_1 X_{1i}(s) + \beta_2 X_{2i}(s)+ \beta_3 X_{3i}(s)+\sigma_y W_y(s).
\end{gather*}
The models for the two missing covariates are given as
\begin{gather*}
X_{2i}(s) \sim N(\mu_{x_{2}i}(s),\tau_{x_2}^{-1}), \hspace{2pt} \mu_{x_{2}i}(s) = \alpha_{20} + \alpha_{21} X_{3i}(s) + \alpha_{22} X_{1i}(s)+ \sigma_{x_2} W_{x_2}(s),\\
X_{1i}(s) \sim N(\mu_{x_{1}i}(s),\tau_{x_1}^{-1}), \hspace{2pt} \mu_{x_{1}i}(s) =  \alpha_{10} + \alpha_{11} X_{3i}(s) + \sigma_{x_1} W_{x_1}(s).
\end{gather*}

The true values of the model parameters are shown in Table \ref{table1}. In order to examine empirical performance of the posterior estimates, several assessment
measures including average bias (Bias), average standard deviations (SD), mean
square error (MSE) and coverage probability (CP) for each parameter are computed.
Taking $\beta_1$ as an example, these measures are given as
\begin{gather*}
\text{Bias} = \frac{1}{T} \sum_{t=1}^{T} (\hat{\beta}_{1t}-\beta_1^0), \hspace{0.2cm}\text{SD} = \frac{1}{T} \sum_{t=1}^{T} \text{sd}(\beta_{1t}),\\
\text{MSE} =  \frac{1}{T} \sum_{t=1}^{T} (\hat{\beta}_{1t}-\beta_1^0)^2, \hspace{0.2cm} \text{CP} = \frac{1}{T}  \sum_{t=1}^{T} \text{1}(\beta_1^0 \in \text{HPD}(\beta_{1t})),
\end{gather*}
where $\beta_1^0$ is the true value of $\beta_1$ and $T$ is the total number of simulated datasets while $\hat{\beta}_{1t}$ is the posterior mean of $\beta_1$. $\text{sd}(\beta_{1t})$ is
the estimated standard deviation of $\beta_1$, and
$\text{HPD}(\beta_{1t})$ is the estimated 95\% highest probability density (HPD) interval of $\beta_1$ computed from the $t$th simulated dataset for $t = 1, \cdots, T$. Bayesian estimates are obtained via JAGS\citep{plummer2003jags} and R\citep{team2013r}. With the
thinning interval to be 20, 5,000 samples are kept for calculation after a burn-in of 10,000 samples. The results of these measures with all records, CC analysis, and model $M_1$ proposed above are shown in Table \ref{table1}. The difference between ``all records", ``CC" and ``$M_1$'' is on the datasets used to fit the proposed model. ``All records" means using the whole dataset before generating the missing ones,  ``CC" means using the datasets excluding the missing records, and ``$M_1$'' means using the datasets with missing values.

\begin{table}[tbp]  
	\centering
	\caption{Simulation results of assessment measures with all records, CC, and model $M_1$} \label{table1}
	\makebox[\linewidth][c]{
		\begin{tabular}{ccccccccccc}
			\hline
			\multirow{2}{1cm}{} & \multirow{2}{1cm}{True value} & \multicolumn{4}{c}{All
				records} & & \multicolumn{4}{c}{CC} \\
			\cline{3-6} \cline{8-11}
			& & Bias & SD & MSE &CP && Bias & SD & MSE  &CP \\
			\hline
			$\beta_0$ & 1 &  -0.0495 & 0.4600  & 0.2140 & 0.97 && 0.2152 & 0.4874  &0.2838
			& 0.94 \\
			$\beta_1$ & 1.5 &  -0.0048 & 0.0710 & 0.0051 & 0.97 && -0.0471 & 0.0975 &
			0.0117 & 0.89 \\
			$\beta_2$ & 1 & 0.0008 & 0.0313 & 0.0010 & 0.95 && -0.0230 & 0.0451 &  0.0026
			& 0.91  \\
			$\beta_3$ & 2 & -0.0014 & 0.0462 & 0.0021 & 0.94 && -0.0279 & 0.0687 &0.0055 & 0.87 \\
			$\tau_y$ & 1 &  -0.0017 & 0.0410 & 0.0017 & 0.94 && 0.0313 & 0.0695 & 0.0058 &
			0.90  \\
			$\alpha_{20}$ & 0 &-0.0375&0.4903&0.2418&0.96&&0.1626&0.4844&0.2611&0.93\\
			$\alpha_{21}$ & 0 &0.0078&0.0460&0.0022&0.96&&-0.0167&0.0585&0.0037&0.93\\
			$\alpha_{22}$ & 2 &-0.0063&0.0314&0.0010&0.95&&-0.0975&0.0498&0.0120&0.49\\
			$\tau_{x_2}$ & 1 &-0.0011&0.0452&0.0020&0.94&&0.0141&0.0573&0.0035&0.97\\
			$ \alpha_{10}$ & 0 &-0.0992&0.4364&0.2003&0.97&&0.5441&0.3190&0.3979&0.86\\
			$ \alpha_{11}$ & 1 &0.0029&0.0318&0.0010&0.94&&-0.2997&0.0702&0.0948&0.00\\
			$\tau_{x_1}$ & 1 &0.0038&0.0422&0.0018&0.94&&0.3301&0.1138&0.1219&0.00\\
			\hline
			\hline
			\multirow{2}{1cm}{} & True &
			\multicolumn{4}{c}{$M_1$}\\
			\cline{3-6}
			&value & Bias & SD & MSE &CP  \\
			\hline
			$\beta_0$ & 1 &  -0.0178 & 0.5091 & 0.2595 & 0.92\\
			$\beta_1$ & 1.5 & 0.0036 & 0.0916 & 0.0084 &0.93 \\
			$\beta_2$ & 1 &  0.0011 & 0.0411 &0.0017 & 0.92 \\
			$\beta_3$ & 2  &-0.0037 & 0.0630 & 0.0040 & 0.92\\
			$\tau_y$ & 1 & 0.0042 & 0.0656 & 0.0043&0.92 \\
			$\alpha_{20}$ & 0 & -0.0386 & 0.4999 & 0.2514& 0.94\\
			$\alpha_{21}$ & 0 & 0.0012 & 0.0549 & 0.0030 &0.95\\
			$\alpha_{22}$ & 2&  0.0023 & 0.0371 & 0.0014 & 0.98\\
			$\tau_{x_2}$ & 1 &-0.0130 & 0.0555 & 0.0033 &0.95\\
			$ \alpha_{10}$ & 0 & 0.0197 & 0.4379 & 0.1921 & 0.94\\
			$ \alpha_{11}$ & 1 &-0.0004 & 0.0334 & 0.0011&0.91\\
			$\tau_{x_1}$ & 1 & 0.0056 & 0.0505 &0.0026 &0.93 \\
			\hline
		\end{tabular}	
	}
\end{table}


From Table \ref{table1}, we can observe that the biases of the posterior estimates under CC are much greater than those under model $M_1$. The 95\% HPD intervals under $M_1$ are larger than those under CC. Thus, $M_1$ is
more preferred than the CC analysis.

In order to assess the performance of the model comparison criteria proposed in
Section \ref{modelassessmentsection}, we set up several alternative models with the same response model
as $M_1$ but with different missing covariate models as follows:
\begin{align*}
M_2: X_{2i}(s) &\sim N(\mu_{x_{2}i}(s),\tau_{x_2}^{-1}), \hspace{2pt} \mu_{x_{2}i}(s) =  \alpha_{20} +
\alpha_{21} X_{3i}(s) + \alpha_{22} X_{1i}(s),\\
&X_{1i}(s) \sim N(\mu_{x_{1}i}(s),\tau_{x_1}^{-1}), \hspace{2pt} \mu_{x_{1}i}(s) =  \alpha_{10} +
\alpha_{11} X_{3i}(s); \\
\\
M_3: X_{2i}(s) &\sim N(\mu_{x_{2}i}(s),\tau_{x_2}^{-1}), \hspace{2pt} \mu_{x_{2}i}(s) = \alpha_{20} +
\alpha_{21} X_{3i}(s) + \alpha_{22} X_{1i}(s),\\
X_{1i}(s) &\sim N(\mu_{x_{1}i}(s),\tau_{x_1}^{-1}), \hspace{2pt} \mu_{x_{1}i}(s) = \alpha_{10} +
\alpha_{11} X_{3i}(s) + \sigma_{x_1} W_{x_1}(s); \\
\\
M_4: X_{2i}(s) \sim &N(\mu_{x_{2}i}(s),\tau_{x_2}^{-1}), \hspace{2pt} \mu_{x_{2}i}(s) = \alpha_{20} +
\alpha_{21} X_{3i}(s) + \alpha_{22} X_{1i}(s)+\sigma_{x_2} W_{x_2}(s),\\
&X_{1i}(s) \sim N(\mu_{x_{1}i}(s),\tau_{x_1}^{-1}), \hspace{2pt} \mu_{x_{1}i}(s) = \alpha_{10} +
\alpha_{11} X_{3i}(s).
\end{align*}

The averages of mDIC and mLPML under these models are shown in Table \ref{table2}. Boxplots of the differences of the mDICs and mLPMLs between each of the missing covariate models $M_2$, $M_3$, and $M_4$ and model $M_1$ are shown in Figure \ref{figure1}.
The boxplots of mDIC and mLPML values for each model are shown in Figure S1 in the supplementary materials.

\begin{table}[tbp]  
	\centering
	\caption{The averages of mDIC and mLPML under $M_1$, $M_2$, $M_3$, and $M_4$} \label{table2}
	\begin{tabular}{ccccc }
		\hline
		& $M_1$ & $M_2$ & $M_3$ & $M_4$ \\
		\hline
		$\text{mDIC}$ & 3151.82 & 3205.51  & 3182.25 & 3171.66
		\\
		$\text{mLPML}$&-1672.62& -1709.00 & -1695.38 & -1683.85
		\\
		\hline
	\end{tabular}
\end{table}

Comparing the mDICs and mLPMLs in Table \ref{table2}, we can see
that model $M_1$ is the best model compared to the other models since it has
the smallest mDIC and the largest mLPML, indicating that these model comparison
criteria perform well in choosing the best model. The simulation results of posterior estimates of parameters in the spatial response model with missing covariate models $M_2$, $M_3$, and $M_4$ are shown in Table S1 in the supplementary materials.

\begin{figure} [h!]  
	\centering
	\includegraphics[height=3in, width=6in]{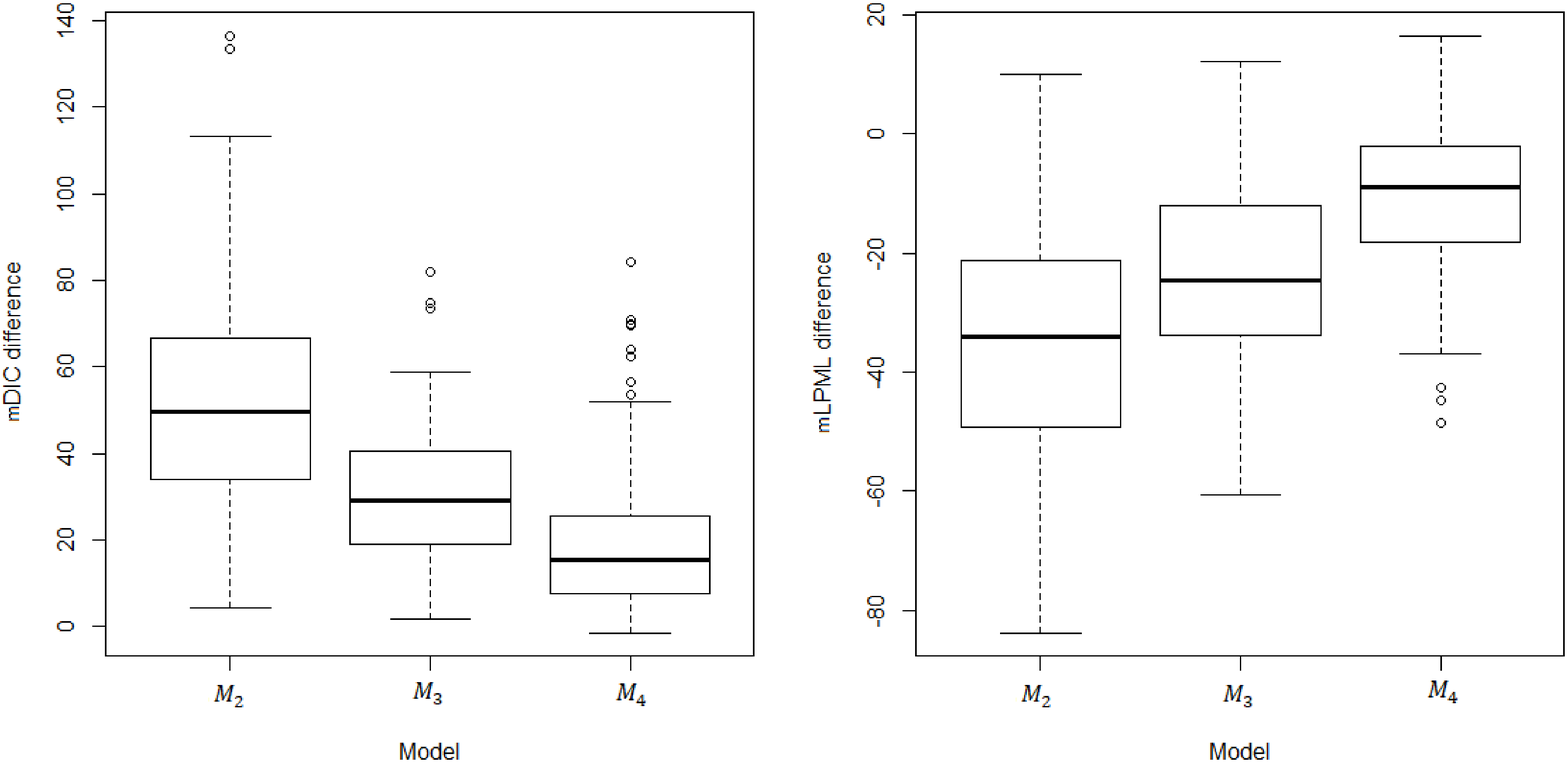}
	\caption{Difference of mDICs and mLPMLs compared to $M_1$} \label{figure1}
\end{figure}

Furthermore, we consider four more estimation models.
Let $M_1^{*}, M^{*}_2, M^{*}_3, M^{*}_4$ denote models of $M_1$, $M_2$, $M_3$, $M_4$ with unknown parameters $\lambda_{y}$ and $\sigma_{y}$. For these two parameters, prior distributions $\lambda_y \sim \text{log-Normal}(0,1)$ and
$\sigma_y^2 \sim \text{half-Normal}(0,0.001^{-1})$ were specified. Model
comparison results of these four models are presented in Table \ref{table3} and
Figure \ref{figure2}.
The boxplots of mDIC and mLPML values for each model are shown in Figure S2 in the supplementary materials.

\begin{table}[tbp]  
	\centering
	\caption{The averages of mDIC and mLPML under $M_1^{*}$, $M_2^{*}$, $M_3^{*}$, and $M_4^{*}$} \label{table3}
	\begin{tabular}{ccccc }
		\hline
		&$M_1^{*}$ & $M_2^{*}$ & $M_3^{*}$ & $M_4^{*}$ \\
		\hline
		$\text{mDIC}$ & 3203.28 & 3271.84  & 3237.75 & 3231.42
		\\
		$\text{mLPML}$&-1710.99 & -1754.99 & -1735.15 &
		-1727.05 \\
		\hline
	\end{tabular}
\end{table}

\begin{figure} [h!]   
	\centering
	\includegraphics[height=3in, width=6in]{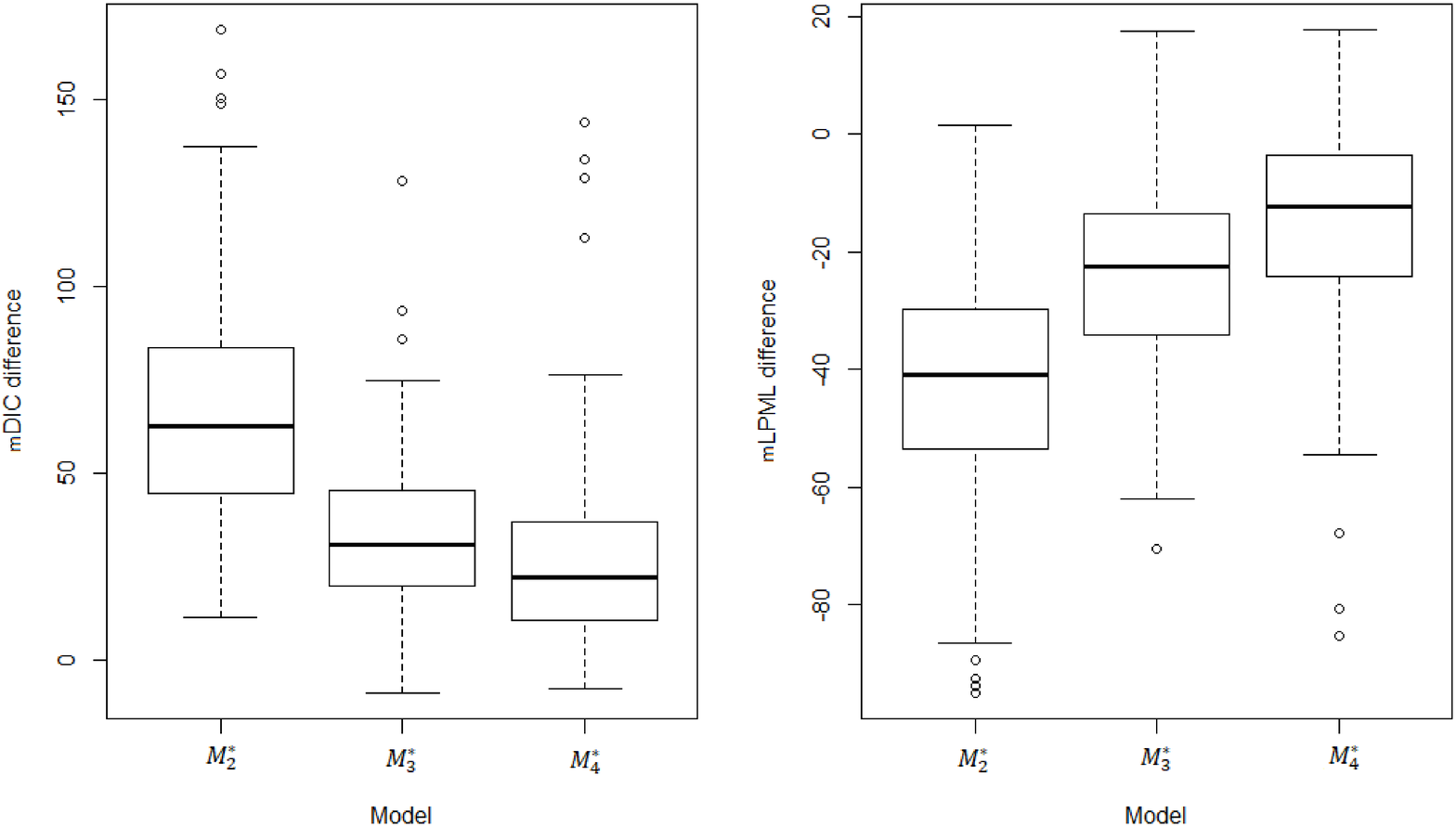}
	\caption{Difference of mDICs and mLPMLs compared to $M_1^{*}$} \label{figure2}
\end{figure}

Similarly, $M^{*}_1$ is the best model chosen by mDIC and mLPML. The results of Bias, SD, MSE and CP for models with all records, CC analysis, and $M_1^{*}$ are shown in Table \ref{table4}. From Table \ref{table4}, similar conclusions can be obtained as Table \ref{table1}. Estimates in CC analysis are biased while CP under model $M_1^{*}$ are generally larger than that of CC. The simulation results of posterior estimates of parameters in the spatial response model with missing covariate models $M_2^{*}$, $M_3^{*}$, and $M_4^{*}$ are shown in Table S2 in the supplementary materials.

\begin{table}[tbp]  
	\renewcommand\arraystretch{0.8}
	\small
	\centering
	\caption{Simulation results of assessment measures with all records, CC, and model $M_1^{*}$}  \label{table4}
	\makebox[\linewidth][c]{
		\begin{tabular}{ccccccccccc}
			\hline
			\multirow{2}{1cm}{} & \multirow{2}{1cm}{True value} & \multicolumn{4}{c}{All
				records*} & & \multicolumn{4}{c}{CC*} \\
			\cline{3-6} \cline{8-11}
			& & Bias & SD & MSE &CP && Bias & SD & MSE  &CP \\
			\hline
			$\beta_0$ &	1&	-0.0553	&0.6026&	0.2888&	0.97 &&	0.3307&	0.5682	&0.3461	&0.92\\
			$\beta_1$ &	1.50	&-0.0112	&0.0712	&0.0044 &0.99	&&-0.0457&	0.0950	&0.0093	&0.96\\
			$\beta_2$ &	1&	-0.0050	&0.0318	&0.0008&	0.96&&	-0.0346&	0.0429&	0.0026	&0.91\\
			$\beta_3$ &	2&	-0.0028	&0.0453	&0.0018&	0.98&&	-0.0138&	0.0611&	0.0032&	0.93\\
			$\sigma_y$ 	 &1.41& 0.0782&	0.3705	&0.1119	&0.96&&	0.0958	&0.3605	&0.0955
			&0.92	\\
			$\text{log}(\lambda_y)$ &1.10&0.5133	&0.9346	&0.7074&1.00&&	0.6476&	0.9300&
			0.8098&	0.93\\
			$\tau_y$ &	1&	0.0006	&0.0452&	0.0019&	0.96&	&0.0242&	0.0631&	0.0050	&0.91\\
			$ \alpha_{20}$ 	&0	&0.0675	&0.4948&	0.2496&	0.91 &&0.3029	&0.5611&	0.2875&
			0.91\\
			$ \alpha_{21}$ &	0&	0.0006&	0.0451&	0.0017&	0.98&&	-0.0186&	0.0609&	0.0036&
			0.95\\
			$\alpha_{22}$& 	2&	0.0002&	0.0317&	0.0009&	0.98&&	-0.1182&	0.0499	&0.0169&
			0.35\\
			$\sigma_{x_2}$ 	& 1.22 &	-0.0079	&0.3075	&0.0535	&0.99&&	0.0822&	0.3338
			&0.0812&	0.98\\
			$\text{log}(\lambda_{x_2})$ &	1.39	&0.5888&	0.9539&	0.8474&	0.95&&0.7008&
			0.9529	&0.8853&	0.93\\
			$\tau_{x_2}$ &	1&	-0.0006&	0.0452&	0.0019	&0.95	&&0.0215	&0.0628&	0.0036&	0.97\\
			$ \alpha_{10}$ &	0&	0.0237&	0.4427&	0.1520	&0.95	&&0.6034	&0.3399	&0.4537	&0.56
			\\
			$ \alpha_{11}$ 	&1	&0.0044	&0.0319	&0.0008	&0.98&&	-0.3297	&0.0441	&0.1154&
			0.00\\
			$\sigma_{x_1}$ 	&1	&-0.0091&	0.2652&	0.0427&	0.99&&	-0.2310&	0.2056&
			0.0824&0.77\\
			$\text{log}(\lambda_{x_1})$ &	1.61&	0.5662&	0.9415&	0.8739	&0.97&&	0.7008&
			0.9529	&0.8853&	0.93\\
			$\tau_{x_1}$& 	1	&-0.0020	&0.0451&	0.0017&	0.97&&	0.3623&	0.0841&	0.1490	&0.01\\
			\hline
			\hline
			\multirow{2}{1cm}{} & True  &
			\multicolumn{4}{c}{$M_1^{*}$}\\
			\cline{3-6}
			& value &Bias & SD & MSE  &CP \\
			\hline
			$\beta_0$ &	1&	0.0891	&0.5836&	0.2598	&0.95\\
			$\beta_1$ &	1.50	&0.0122	&0.0919&	0.0065	&0.98\\
			$\beta_2$ &	1&	-0.0048	&0.0412&	0.0013&	0.97\\
			$\beta_3$ &	2&	0.0099	&0.0570&	0.0027	&0.98\\
			$\sigma_y$ 	 &1.41&0.0814&	0.3731&	0.1170	&0.94\\
			$\text{log}(\lambda_y)$ &1.10& 0.5852	&0.9262&0.8077 &	0.98\\
			$\tau_y$ &	1&	-0.0055	&0.0599	&0.0036	&0.94\\
			$ \alpha_{20}$ 	&0	&0.0503	&0.5429&	0.2184	&0.93\\
			$ \alpha_{21}$ &	0&	0.0052&	0.0574&	0.0027&	0.96\\
			$\alpha_{22}$& 	2&	-0.0068&	0.0451&	0.0016	&0.98\\
			$\sigma_{x_2}$ 	& 1.22 &	0.0585	&0.3330&	0.0911	&0.96\\
			$\text{log}(\lambda_{x_2})$ &1.39&	0.6294&	0.9473&0.8727&0.91\\
			$\tau_{x_2}$ &	1&	-0.0082	&0.0603&	0.0030	&0.98\\
			$ \alpha_{10}$ &	0&-0.0279	&0.4938&	0.2202&	0.90\\
			$ \alpha_{11}$ 	&1	&	0.0016&0.0341&	0.0013	&0.93\\
			$\sigma_{x_1}$ 	&1	&	0.0075	&0.2774&	0.0513&	0.96\\
			$\text{log}(\lambda_{x_1})$ &	1.61&0.6479&	0.9507&	0.8914&	0.97\\
			$\tau_{x_1}$& 	1	&0.0075&	0.0515&	0.0026&	0.96\\
			\hline
		\end{tabular}	
	}
\end{table}

\section{Application to Spatial Health and Nutrition Survey Data} \label{appsection}
In this section, the proposed Bayesian hierarchical spatial model and model
comparison criteria are applied to analyze the CHNS 2011 survey data described
in Section \ref{datasection}.
\subsection{Real Data Model}
For the spatial positive continuous response variable $hincome$, the spatial
regression model proposed in Section \ref{resmodelsection} are built for its logarithm form.
Covariate vector $\bm{X}$ involves five individual covariates including
\text{log}(\textit{indwage}), \textit{age}, \textit{urban}, \text{log}(\textit{WThour}) and \textit{hhsize} and a
province-level covariate \textit{GDP}. The $(s,s^{'})$th entry of
$H(\lambda_y)$ is $\exp(-d_{ss^{'}}/\lambda_y)$, where $d_{ss^{'}}$ is the
distance between location $s$ and location $s^{'}$.

For the individual-level covariates, two of them, \textit{indwage} and \textit{WThour}, are
missing. In order to take account of different spatial structures in the
missing covariates, we consider four different missing covariate models in our
study.

Denote $\text{log}(hincome)$ as $Y$, $\text{log}(WThour)$ as $X_1$, $\text{log}(indwage)$ as $X_2$, $GDP$ as $X_3$, $age$ as $X_4$, $urban$ as $X_5$, and $hhsize$ as $X_6$. We first consider the following model $M_1^{real}$ for the data.
The spatial regression model for the response variable is:
\begin{gather*}
Y_i(s) = \beta_0+ \beta_1 X_{1i}(s) + \beta_2 X_{2i}(s) + \beta_3 X_{3i}(s) + \beta_4 X_{4i}(s)  +\beta_5 X_{5i}(s) +\beta_6 X_{6i}(s) + \sigma_y W_y(s) + \epsilon_{yi}(s),\\
\epsilon_{yi}(s) \sim N(0,\tau_y^{-1}),\bm{W}_y \sim \text{MVN}(\bm{0}, H(\lambda_y)).
\end{gather*}
The missing covariate model is:
\begin{gather*}
X_{2i}(s) =  \alpha_{20}+
\alpha_{21} X_{1i}(s) + \alpha_{22} X_{3i}(s) + \alpha_{23} X_{4i}(s)  + \alpha_{24} X_{5i}(s) + \sigma_{x_2}
W_{x_2}(s)+\epsilon_{x_2 i}(s),\\
\epsilon_{x_2 i}(s)\sim N(0, \tau_{x_2}^{-1}), \bm{W}_{x_2} \sim \text{MVN}(\bm{0},
H(\lambda_{x_2}));\\
X_{1i}(s)  = \alpha_{10}+ \alpha_{11} X_{3i}(s)  +  \alpha_{12} X_{4i}(s)  +
\alpha_{13}X_{5i}(s)  +  \alpha_{14} X_{6i}(s)  + \sigma_{x_1} W_{x_1}(s)+\epsilon_{x_1 i}(s),\\
\epsilon_{x_1 i}(s) \sim N(0, \tau_{x_1}^{-1}), \bm{W}_{x_1} \sim
\text{MVN}(\bm{0},H(\lambda_{x_1})).
\end{gather*}
We also consider another three alternative models with the same response model
as $M_1^{real}$ but with different missing covariate distributions as follows:
\begin{gather*}
M_2^{real}: X_{2i}(s)  =\alpha_{20}+
\alpha_{21} X_{1i}(s)  + \alpha_{22} X_{3i}(s)  + \alpha_{23} X_{4i}(s)   + \alpha_{24} X_{5i}(s)
+\epsilon_{x_2 i}(s),\\
X_{1i}(s)  = \alpha_{10}+ \alpha_{11} X_{3i}(s)  +  \alpha_{12} X_{4i}(s)  +
\alpha_{13}X_{5i}(s)  +  \alpha_{14} X_{6i}(s)   +\epsilon_{x_1 i}(s),\\
\epsilon_{x_2 i}(s) \sim N(0, \tau_{x_2}^{-1}), \hspace{0.2cm} \epsilon_{x_1 i}(s)\sim N(0, \tau_{x_1}^{-1}).
\end{gather*}
\begin{gather*}
M_3^{real}: X_{2i}(s)  =\alpha_{20}+
\alpha_{21} X_{1i}(s)  + \alpha_{22} X_{3i}(s)  + \alpha_{23} X_{4i}(s)   + \alpha_{24} X_{5i}(s)   + \sigma_{x_2} W_{x_2}(s)+\epsilon_{x_2 i}(s),\\
X_{1i}(s)  =  \alpha_{10}+ \alpha_{11}X_{3i}(s)   +  \alpha_{12} X_{4i}(s) +
\alpha_{13}X_{5i}(s) +  \alpha_{14} X_{6i}(s)  +\epsilon_{x_1 i}(s),\\
\epsilon_{x_2 i}(s) \sim N(0, \tau_{x_2}^{-1}), \bm{W}_{x_2} \sim \text{MVN}(\bm{0},
H(\lambda_{x_2})), \hspace{0.2cm} \epsilon_{x_1 i}(s) \sim N(0, \tau_{x_1}^{-1}).
\end{gather*}
\begin{gather*}
M_4^{real}: X_{2i}(s)  =\alpha_{20}+
\alpha_{21} X_{1i}(s)  + \alpha_{22} X_{3i}(s)  + \alpha_{23} X_{4i}(s)   + \alpha_{24} X_{5i}(s)
+\epsilon_{x_2 i}(s),\\
X_{1i}(s)  =  \alpha_{10}+ \alpha_{11} X_{3i}(s)  +  \alpha_{12} X_{4i}(s)  +
\alpha_{13}X_{5i}(s)  +  \alpha_{14} X_{6i}(s)  +\sigma_{x_1}
W_{x_1}(s)+\epsilon_{x_1 i}(s),\\
\epsilon_{x_2 i}(s)\sim N(0, \tau_{x_2}^{-1}), \hspace{0.2cm} \epsilon_{x_1 i}(s)\sim N(0, \tau_{x_1}^{-1}), \bm{W}_{x_1} \sim \text{MVN}(\bm{0},
H(\lambda_{x_1})).
\end{gather*}

Assume $\bm{R}_1$ and $\bm{R}_2$ represent the missing
indicators of covariates $WThour$ and $indwage$ respectively, where $R_{\ell i}(s)=1$
denotes missing records and $R_{\ell i}(s)=0$ denotes observed ones ($\ell=1,2$) at location $s$.
For each of the above four models, the following MAR model is assumed for the missing data mechanism:
\begin{gather*}
M_R^{m_1}: R_{1i}(s) \sim \text{Bernoulli}(p_{1i}(s)), \hspace{3pt} R_{2i}(s) \sim \text{Bernoulli}(p_{2i}(s)),\\
\text{logit}(p_{1i}(s)) = \phi_{10} +  \phi_{11} X_{3i}(s)   + \phi_{12} X_{4i}(s)  +
\phi_{13} X_{5i}(s) +  \phi_{14} X_{6i}(s)  + \phi_{15} Y_{i}(s) ;\\
\text{logit}(p_{2i}(s)) =  \phi_{20}+ \phi_{21} X_{3i}(s)   +  \phi_{22} X_{4i}(s)  +
\phi_{23} X_{5i}(s)  +  \phi_{24} X_{6i}(s)  + \phi_{25} Y_i(s).
\end{gather*}

The same prior distributions described in Section \ref{priorsection} were used in these four
competitive models along with model $M_R^{m_1}$ for the missing data mechanism. mDIC and mLPML values under models $M_1^{real}$ to $M_4^{real}$ are calculated via JAGS and R. With the thinning interval to be 25, 8,000 samples
are kept for calculation after a burn-in of 150,000 samples.
The convergence of the MCMC sampling algorithm is checked using several diagnostic procedures discussed in \citet{cowles1996markov} and \citet{chen2000monte}. For example, the traceplots of the parameters under model $M_1^{real}$ shown in Figure S3 demonstrate good mixing of MCMC chains.

\subsection{Real Data Results}
Table \ref{table7} shows the values of mDIC and mLPML under the four models for the CHNS 2011 survey data. From Table \ref{table7}, we choose model $M_1^{real}$ since it has the smallest mDIC and the largest mLPML among these models.
The posterior estimates of the parameters under model $M_1^{real}$ and the results of CC estimation are given in Table \ref{table8}. From Table \ref{table8}, we can observe that covariates \textit{GDP},
\textit{indwage}, \textit{age}, \textit{urban}, and \textit{hhsize}, have significant positive impact on the
household income. The household income has spatial correlation among different
provinces. For missing covariate \textit{indwage}, both \textit{GDP}, \textit{age}, \textit{WThour}, and
\textit{urban} have significant impact on it. \textit{age}, and \textit{urban} can help explain the missing
covariate \textit{WThour}. These two missing covariates also have spatial correlation
among different provinces like the household income. The posterior estimates of parameters under other models, namely, $M_2^{real}$, $M_3^{real}$, and $M_4^{real}$, are shown in Table S3 in the supplementary materials. Since under model $M_R^{m_1}$ (MAR), $\bm{\phi}$ is independent of
the other parameters {\it a posteriori}, the posterior estimates of $\bm{\phi}$ remain the same no matter which of  models $M_1^{real}$ to $M_4^{real}$ is used to fit  \textit{hincome}, \textit{indwage} and \textit{WThour}.
These estimates are reported in Table \ref{table9}. We see from Table \ref{table9} that  the 95\% HPD intervals for $\phi_{11}$,
$\phi_{12}$, $\phi_{13}$, $\phi_{14}$, $\phi_{15}$, $\phi_{22}$, and  $\phi_{25}$ do not contain zero, implying that the missingness mechanism is not missing completely at random (MCAR).

\begin{table}[tbp]   
	\centering
	\caption{Results of model comparison for the CHNS 2011 survey data} \label{table7}
	\begin{tabular}{ccccc}
		\hline
		Model & $M_1^{real}$ & $M_2^{real}$ & $M_3^{real}$ & $M_4^{real}$\\
		\hline
		$\text{mDIC}$ & 8968.30 & 8982.39 & 8971.00 & 8970.43\\
		$\text{mLPML}$ & -4541.60 & -4558.69  & -4546.51
		&-4543.62 \\
		\hline
	\end{tabular}	
\end{table}

\begin{table}[tbp]
	\centering
	\small
	\caption{Posterior estimates under CC and $M_1^{real}$ for the CHNS 2011 survey data} \label{table8}
	\makebox[\linewidth][c]{
		\begin{tabular}{cccccccc}
			\hline
			\multirow{2}{1cm}{} & \multicolumn{3}{c}{CC} & &
			\multicolumn{3}{c}{$M_1^{real}$}  \\
			\cline{2-4} \cline{6-8}
			Parameters & Mean & SD & 95\% HPD interval & & Mean & SD & 95\%  HPD interval \\
			\hline
			$\beta_0$ & 6.4474 & 0.1280 & (6.2027, 6.6957) && 6.9041&0.1008&(6.7067,
			7.1053)\\
			$\beta_1$ & -0.0017 & 0.0143 & (-0.0295, 0.0256) && 0.0072&0.0134&(-0.0188,
			0.0336)\\
			$\beta_2$ & 0.3885 & 0.0115 & (0.3663, 0.4108) &&
			0.3271&0.0096&(0.3081,0.3455)\\
			$\beta_3$ & 0.1351 & 0.0558 & (0.0295, 0.2518) && 0.1607&0.0438&(0.0730,
			0.2482)\\
			$\beta_4$ & 0.0676 & 0.0144 & (0.0396, 0.0965) &&0.1652
			&0.0135&(0.1392,0.1914)\\
			$\beta_5$ & 0.0595 & 0.0309 & (0.0003, 0.1215) &&
			0.3003&0.0249&(0.2518,0.3499)\\
			$\beta_6$ & 0.1377 & 0.0108 & (0.1173, 0.1597) &&
			0.1491&0.0082&(0.1329,0.1654)\\
			$\tau_y$ & 2.8863 & 0.0892 & (2.7145, 3.0665) && 2.4762&0.0636&(2.3525,
			2.6004)\\
			$\text{log}(\lambda_y)$ & -0.1595 & 0.8870 & (-1.9897, 1.4398) &&-0.0778
			&0.9261&(-1.9814, 1.6019)\\
			$\sigma_y$ & 0.1635 & 0.0513 & (0.0890, 0.2916) &&0.1383 &0.0429&(0.0691,
			0.2413)\\
			$\alpha_{20}$ &8.8296&0.1781&(8.4922, 9.1953)&& 8.3155 & 0.1693 & (7.9573,
			8.6412) \\
			$\alpha_{21}$ &0.1548&0.0267&(0.1010, 0.2082)&& 0.1973 & 0.0303 & (0.1368,
			0.2557) \\
			$\alpha_{22}$ &0.2361&0.1133&(0.0068, 0.4672)&& 0.2656 & 0.0918 & (0.0709,
			0.4396)\\
			$\alpha_{23}$ &-0.4240&0.0265&(-0.4758, -0.3728)&& -0.6901 & 0.0267 & (-0.7436,
			-0.6395) \\
			$\alpha_{24}$ &0.6091&0.0550&(0.5008, 0.7150)&& 0.3911 & 0.0505 & (0.2915,
			0.4891)\\
			$\tau_{x_2}$ &0.7979&0.0243&(0.7504, 0.8462)&& 0.6730 & 0.0198 & (0.6362,
			0.7127) \\
			$\text{log}(\lambda_{x_2})$ &0.1633&1.1194&(-1.9645, 2.3496)&& 0.6740 & 1.3232
			& (-1.7436, 3.2073) \\
			$\sigma_{x_2}$ &0.3711&0.1282&(0.2043, 0.7056)&& 0.2968  &0.1419 & (0.1352,
			0.7163) \\
			$\alpha_{10}$ &3.4173&0.0906&(3.2414, 3.5955)&& 3.1448 & 0.2728 & (2.4942,
			3.7548) \\
			$\alpha_{11}$ &0.0478&0.0619&(-0.0804, 0.1730)&& 0.0618 & 0.1660 & (-0.2334,
			0.4624)\\
			$\alpha_{12}$ &-0.1014&0.0206&(-0.1407, -0.0597)&& -0.1276 & 0.0240 & (-0.1755,
			-0.0803) \\
			$\alpha_{13}$ &0.3210&0.0452&(0.2309, 0.4084)&& 0.4926 & 0.0450 & (0.4034,
			0.5791) \\
			$\alpha_{14}$ &-0.0333&0.0165&(-0.0663, -0.0004)&& -0.0202 & 0.0145 & (-0.0495,
			0.0088) \\
			$\tau_{x_1}$ &1.2309&0.0375&(1.1590, 1.3053)&& 0.9315 & 0.0237 & (0.8857,
			0.9775)\\
			$\text{log}(\lambda_{x_1})$ & -0.1166&0.9361&(-1.9674, 1.6887)&& 0.7219 &
			1.2951 & (-1.7500, 3.1023) \\
			$\sigma_{x_1}$ &0.1828&0.0629&(0.0949, 0.3384)&& 0.4659  &0.1581 & (0.2450,
			0.8626) \\
			\hline
		\end{tabular}	
	}
\end{table}

\begin{table}[tbp]
	\centering
	\small
	\caption{Posterior estimates of the missingness mechanism model $M_R^{m_1}$} \label{table9}
	\makebox[\linewidth][c]{
		\begin{tabular}{ccccccccc}
			\hline
			Parameters & Mean & SD & 95\% HPD interval && Parameters & Mean & SD & 95\% HPD interval\\
			\cline{1-4} \cline{6-9}
            $\phi_{10}$ & -4.3012&0.6841&(-5.6261,-2.9422)
			          && $\phi_{20}$& 3.7527&0.4645&(2.8064,4.6907)  \\
            $\phi_{11}$ & 0.2065&0.05&(0.1109,0.3063)
                      && $\phi_{21}$ & -0.0315&0.0378&(-0.1046,0.0435)  \\
			$\phi_{12}$ & 2.0856&0.0734&(1.9445,2.2289) &&
                 $\phi_{22}$ &  0.8108&0.0376&(0.739,0.886)  \\
			$\phi_{13}$ & 2.0796&0.1118&(1.8607,2.2974)  &&
			$\phi_{23}$ &  -0.1247&0.0743&(-0.2702,0.0226) \\
			$\phi_{14}$ & -0.1776&0.0362&(-0.2484,-0.1073)  &&
			$\phi_{24}$ & 0.0121&0.0255&(-0.0377,0.0622) \\
			$\phi_{15}$ & 0.2036&0.0682&(0.0679,0.3353) &&
			$\phi_{25}$ & -0.3768&0.0458&(-0.4691,-0.2824) \\
			\hline
		\end{tabular}	
	}
\end{table}

In order to see whether the posterior estimates will differ with different spatial structures, we also consider another commonly used spatial structure, the conditional autoregressive (CAR) structure, in our analysis. With a similar form with model $M_1^{real}$, we assume that both $\bm{W}_y$, $\bm{W}_{x_2}$ and $\bm{W}_{x_1}$ follow a CAR structure $\text{MVN}(\bm{0}, \Sigma_w)$ with $\Sigma_w = (I - \lambda D)^{-1}$, where $I = \text{diag}(1)$ and $D$ is the adjacent matrix of the 12 locations. The mDIC and mLPML values of this model with CAR structures are 8966.27 and -4540.83, which are quite close to those under model $M_1^{real}$. The posterior estimates under the model with CAR structure are also similar, which are shown in Table S4 in the supplementary materials.

We also carry out a sensitivity analysis on specification of the models for missing data mechanism.
In addition to model $M_R^{m_1}$ (MAR), we further consider a non-ignorable model for the two missing covariates, given by
\begin{equation*}
\begin{split}
	&M_R^{m_2}: R_{1i}(s) \sim \text{Bernoulli}(p_{1i}(s)), \hspace{3pt} R_{2i}(s) \sim \text{Bernoulli}(p_{2i}(s)),\\
&\text{logit}(p_{1i}(s)) = \phi_{10} +  \phi_{11} X_{3i}(s)   + \phi_{12} X_{4i}(s)  +
\phi_{13} X_{5i}(s)  \\
&\qquad \qquad\qquad+  \phi_{14} X_{6i}(s)  + \phi_{15} Y_{i}(s) + \phi_{16} W_{x_2}(s)  + \phi_{17}
W_{x_1}(s) ;\\
&\text{logit}(p_{2i}(s)) =  \phi_{20}+ \phi_{21} X_{3i}(s) 
+  \phi_{22} X_{4i}(s) \\
& \qquad\qquad\qquad+
\phi_{23} X_{5i}(s)  +  \phi_{24} X_{6i}(s)   + \phi_{25} Y_i(s) + \phi_{26} W_{x_2}(s)  + \phi_{27} W_{x_1}(s).
\end{split}
\end{equation*}

With the thinning interval to be 25, 8,000 samples
are kept for calculation after a burn-in of 150,000 samples using JAGS and R.
With the same response model and missing covariates model as model $M_1^{real}$, we fit the one with a different missing mechanism model $M_R^{m_2}$. Posterior estimates under this model is shown in Table S5 
in the supplementary materials. By comparing the estimates under this model and $M_1^{real}$, we can see that the estimates of parameters in the response model and the missing covariate distribution are quite similar, so $M_1^{real}$ with a MAR assumption is a relatively simple model to achieve our goal of analysis.

We calculate the mDIC for the response model as well as $\text{DIC}(R)$
for the missingness mechanism model alone. $\text{DIC}(R)$ is defined with
$\text{Dev}(\bar{\theta}) = -2 \text{log}f(\bm{R}_1,\bm{R}_2|\bm{\phi},\bm{X})$, where $\bm{X}$
denotes the covariates included in the missingness mechanism models.

The mDIC values under models $M_1^{real}$ to $M_4^{real}$ are 8968.30, 8982.39, 8971.00, and 8970.43, respectively,
under the MAR missingness model $M_R^{m_1}$, while  these mDIC values are 8966.80, 8979.23, 8970.98, and 8969.02, respectively,
under model $M_R^{m_2}$.
For models $M_1^{real}$ to $M_4^{real}$ with the MAR missingness model $M_R^{m_1}$, the $\text{DIC}(R)$ values are 8122.12, 8121.89, 8122.19, and 8122.03.
For models with the missingness model $M_R^{m_2}$,  the $\text{DIC}(R)$ values are 7963.12, 7965.87, 7963.98, and 7963.75, corresponding to models $M_1^{real}$ to $M_4^{real}$, respectively.
These results show that models with $M_R^{m_2}$ as missingness model have lower mDIC values and $\text{DIC}(R)$ values, therefore, we can conclude that the
missingness mechanism model $M_R^{m_2}$ is preferred.

The posterior estimates of $\bm{\phi}$ under model $M_R^{m_2}$ are given in Table S6 in the supplemental materials. For the chosen model $M_1^{real}$ with missingness model $M_R^{m_2}$, we can see that $age$, $hhsize$ and the spatial effects $\bm{W}_{x_2}$ have a significant positive effect on the missingness of covariate $indwage$, while the response variable has a significant negative
effect on the missingness of covariate $indwage$. It means that older people, people with a higher household income, and people who have a larger family are prone to
reject to report their wages. For the missingness of covariate $WThour$, both the two spatial effects have the significant positive impact. Older people, people living in the
urban area, people who have a smaller family and people with a higher household income tend to reject to report their working
hours in this analysis. In addition, the coefficients of the spatial effects, $\phi_{16}$, $\phi_{17}$, and $\phi_{26}$, are significant,
 meaning that the missingness of the missing covariates does depend on the spatial random effects.

\section{Discussion} \label{discussion}
In this paper, a Bayesian hierarchical spatial model is constructed for spatial
data with missing covariates. In addition to a Gaussian stationary spatial
process model for the continuous spatial response, missing covariate models
with spatial random effects are built for the missing covariates. In our
method, missingness mechanisms for the missing covariates are restricted to be
MAR, which may not be suitable in practice. MNAR is more common in reality and
may introduce much more complexity in analysis. Future study can be focus on
extending the missingness mechanism to be MNAR, and missingness mechanism
models should be built to test the assumptions of missingness mechanisms.
Additionally, in our method, a spatial model is built for the continuous
response variable, which can be extended to variables of other data types, such
as categorical responses. In the real data analysis, we also fit the models using the conditional autoregressive (CAR) spatial random effects in both the response model and the missing covariate models.
From the results in the supplementary materials, we find that the Gaussian random effects and the CAR random effects yield nearly the same estimation results. 
In the future, we can introduce missing covariates
model to autologistic model which is universally used for spatial binary
data. Furthermore, dealing with spatial effects and missing variables
simultaneously complicates the implementation of MCMC sampling algorithms, so it is also necessary to develop efficient algorithms and software to speed up convergence of MCMC sampling. One limitation of the data we analyzed is the lack of detailed address information for households. In this study, we just emphasized on the spatial dependent structure at the province level. Considering both between-province dependency and within-province dependency is an area devoted for future research.

\section*{Acknowledgements}
We would like to thank the Editor-in-Chief and the Referee for their very helpful comments and suggestions, which helped us further improve the paper.
Dr. Chen's research was partially supported by NIH grants \#GM70335 and  \#P01CA142538. Dr. Hu's research was supported by Dean's office of the College of Liberal Arts and Sciences at University of Connecticut. Dr. Ma's research was supported by Project of Educational Commission of  Guangdong Province of China \#2019WQNCX104.
\section*{Data Availability Statement}
The data that support the findings of this study are available from the corresponding author upon reasonable request.
\section*{Supporting Information}
Additional figures and tables for this article are available online, including boxplots of mDIC and mLPML under models in the simulation study, trace plots of parameters under $M_1^{real}$, simulation results for models $M_2$, $M_3$, $M_4$, $M_2^*$, $M_3^*$ and $M_4^*$, posterior estimates under model $M_2^{real}$, $M_3^{real}$, $M_4^{real}$, the model with CAR structure and models with missingness model $M_R^{m2}$ in real data analysis.

\end{document}